\documentclass[letter]{aa}

\usepackage{graphicx}
\usepackage{txfonts}
\usepackage{stmaryrd}
\usepackage{upgreek}
\usepackage[colorlinks,citecolor=blue,urlcolor=blue,filecolor=blue,linkcolor=blue]{hyperref}

\defcitealias{itu_m1583_1}{Rec.~ITU-R~M.1583-1}
\defcitealias{itu_ra769_2}{Rec.~ITU-R~RA.769-2}
\defcitealias{itu_s1586_1}{Rec.~ITU-R~S.1586-1}
\defcitealias{cispr32}{CISPR-32}
\defcitealias{en55032}{EN\,55032}
\defcitealias{radioregs}{Radio Regulations}

\begin{document}

   \title{Bright unintended electromagnetic radiation from second-generation Starlink satellites}

   \author{
     C.\,G.\,Bassa\inst{\ref{astron}}\fnmsep\thanks{Member of the IAU Centre for the Protection of the Dark and Quiet Sky from Satellite Constellation Interference (IAU CPS).} 
     \and
     F.\,Di\,Vruno\inst{\ref{skao},\ref{craf}}\fnmsep$^\star$
     \and  
     B.\,Winkel\inst{\ref{mpifr},\ref{craf}}\fnmsep$^\star$
     \and
     G.\,I.\,G.\,J\'{o}zsa\inst{\ref{mpifr},\ref{rhodes},\ref{craf}}\fnmsep$^\star$
     \and
     M.\,A.\,Brentjens\inst{\ref{astron}}
     \and
     X.\,Zhang\inst{\ref{lesia}}
   }

   \institute{ASTRON, Netherlands Institute for Radio Astronomy, Oude
     Hoogeveensedijk 4, 7991 PD Dwingeloo, The
     Netherlands\\ \email{bassa@astron.nl}
     \label{astron}
     \and
     Square Kilometre Array Observatory, Lower Withington,
     Macclesfield, Cheshire, SK11 9FT, United Kingdom\label{skao}
     \and
     European Science Foundation, Committee on Radio Astronomy
     Frequencies, 1, quai Lezay Marnésia BP 90015, F-67080 Strasbourg
     Cedex, France\label{craf}
     \and
     Max-Planck-Institut f\"ur Radioastronomie, Auf dem H\"ugel
     69, 53121 Bonn, Germany\label{mpifr}
     \and
     Department of Physics and Electronics, Rhodes University, PO
     Box 94, Makhanda, 6140, South Africa\label{rhodes}
     \and
     LESIA, Observatoire de Paris, Université PSL, CNRS, 5 place Jules Janssen, 92195 Meudon, France\label{lesia}
   }

   \date{Received 9 August 2024; accepted 29 August 2024}

   \abstract{We report on the detection of unintended electromagnetic radiation (UEMR) from the second-generation of Starlink satellites. Observations with the LOFAR radio telescope between 10 to 88\,MHz and 110 to 188\,MHz  show  broadband emission covering the frequency ranges from 40 to 70\,MHz and 110 to 188\,MHz from the v2-Mini and v2-Mini Direct-to-Cell Starlink satellites. The spectral power flux density of this broadband UEMR varies from satellite to satellite, with values ranging from 15\,Jy to 1300\,Jy, between 56 and 66\,MHz, and from 2 to 100\,Jy over two distinct 8\,MHz frequency ranges centered at 120 and 161\,MHz. We compared the detected power flux densities of this UEMR to that emitted by the first generation v1.0 and v1.5 Starlink satellites. When correcting for the observed satellite distances, we find that the second-generation satellites emit UEMR that is  up to a factor of 32 stronger  compared to the first generation. The calculated electric field strengths of the detected UEMR exceed typical electromagnetic compatibility standards used for commercial electronic devices as well as recommended emission thresholds from the Radiocommunication Sector of the International Telecommunications Union (ITU-R) aimed at protecting the 150.05--153\,MHz frequency range allocated to radio astronomy. We characterize the properties of the detected UEMR with the aim of assisting the satellite operator with the identification of the cause of the UEMR.}


   \maketitle

\section{Introduction}   
With the miniaturization of satellites and the rapid commercialization of spaceflight, the number of satellites in orbit around the Earth has dramatically increased since around 2016 \citep[see][]{mcdowell2020}. Since then, several commercial companies have started to mass produce and launch large numbers of satellites to offer various communication services, primarily broadband internet and mobile connectivity. Satellites in such constellations are typically launched into shells of specific orbital altitude and orbital inclination to optimize coverage for specific geographic latitudes.

This growth in the number of satellites has worrisome implications for astronomy \citep{walker+2020_dqs1,walker+2020,walker+2021_dqs2} as the probability that satellites are passing through the fields of view of ground-based telescopes (as well as space-based)  is drastically increasing, and sunlight reflected from these satellites, as well as radio signals emitted by them,  are detectable by astronomical instruments 
\citep[e.g.,][]{tyson+2020,michalowski+2021,mroz+2022,kruk+2023}. Assessing the impact of these satellite constellations on different astronomical observatories and science cases is an increasingly important topic of recent research \citep[e.g.,][]{green+2022,bassa+2022,barentine+2023,lang+2023,kovalev+2023,hainaut+2024}.

For radio astronomy, the use of the radio spectrum is regulated by the Radiocommunication Sector of the International Telecommunication Union (ITU-R), which publishes the relevant international treaty in the form of the \citetalias{radioregs}. These regulations cover the intentional use of the radio spectrum for different applications (or services) such as communication, remote sensing, navigation, as well as astronomy. It considers wanted and unwanted emission, for example that due to  out-of-band emission from spectral side-lobes. In particular, the ITU-R allocates several frequency ranges to the radio astronomy service. \citetalias{itu_ra769_2} provides thresholds on the received power (or power flux densities) that must not be exceeded by other active radio services in these bands. As these bands focus on spectral lines affected by Galactic Doppler shifts, the frequency ranges are relatively narrow. 

In \citet{divruno+2023} we introduced the concept of unintended electromagnetic radiation (UEMR), referring to any electromagnetic radiation that is radiated by (or leaking from) electrical devices and systems on board satellites, and not necessarily related to the generation and transmission of wanted electromagnetic radiation from antennas used for  communication, for example. Through simulations of the aggregate effect of UEMR emitted by satellites in various large satellite constellations, we found that the radiation levels required to comply with ITU-R specified interference thresholds for frequency ranges assigned to radio astronomy, would be quite constraining when compared to typical electromagnetic compatibility standards used for commercial devices on Earth. Unfortunately, the detected levels of both narrowband and broadband UEMR at frequencies between 110 and 188\,MHz of dozens of satellites belonging to the SpaceX Starlink constellation showed emissions above these calculated thresholds \citep{divruno+2023}. This detection of UEMR from the Starlink constellation has since been independently confirmed by \citet{grigg+2023}.

In \citet{divruno+2023} we argued that while narrowband UEMR detected at 143.050\,MHz could be attributed to reflections from the French GRAVES Space Surveillance Radar, other narrow- and broadband emissions were intrinsically radiated by the satellites. Doppler analysis of the narrowband emissions provides further evidence (Bassa et al., in prep.) of this intrinsic origin. Triggered by the initial detections of satellite UEMR we have initiated an observing program to investigate and characterize UEMR from satellites within different satellite constellations (e.g.,\ OneWeb, IRIDIUM Next, Swarm, Planet Labs, BlueWalker), and different hardware versions of satellites within a constellation (Starlink). As part of these observations, and those triggered by a request (C.\ Lonsdale, priv.\ comm.) to confirm circumstantial evidence for possible UEMR detected in the EDGES epoch-of-reionization experiment \citep{bowman+2008}, we report on the detection of bright and broadband UEMR from satellites of the second generation of Starlink.

\section{Observations and analysis}
\subsection{Observations}
The observations reported here closely follow the observational setup used and detailed in \citet{divruno+2023}, where satellites are detected by letting them pass through the telescope beam pattern. Two one-hour observations were obtained using the central six stations of the LOFAR radio telescope \citep{vanhaarlem+2013} in the Netherlands on July 19, 2024, one covering frequencies from 10 to 88\,MHz using the low-band antennas (LBAs) in the LBA\_OUTER configuration, the other 110 to 188\,MHz with the high-band antennas (HBAs). Signals from these stations were coherently added in the \texttt{COBALT} beamformer \citep{broekema+2018} to form 91 tied-array beams (TABs), tiling out the primary station beam in hexagonal rings with separations near the TAB full width at half maximum (FWHM) of $42\arcmin$ for the LBA observation, and $24\arcmin$ for the HBA observation. For each tied-array beam, dynamic spectra with total intensity (Stokes I) were recorded at 41.94\,ms time resolution and 12.2\,kHz frequency resolution. To minimize the distance between the telescope and satellites passing through the beam pattern, the TABs for both observations   tracked equatorial positions, which culminated near zenith (maximum elevation $87\fdg5$) half-way through the one-hour integration. 

\subsection{Analysis}
The dynamic spectra were analyzed and searched for the presence of satellite UEMR using an adapted version of the method outlined in \citet{divruno+2023}. First, we retrieved orbital elements in the form of two-line element sets (TLEs), from publicly available catalogs. These orbital elements are derived from observations by the United States Space Force (USSF).\footnote{Distributed through \url{www.space-track.org}.} This approach allows searching for the UEMR of any satellite in the USSF catalog, whereas in \citet{divruno+2023} we used ephemerides provided by SpaceX that are only available for Starlink satellites. Using these orbital elements, we used the \texttt{Skyfield} software to compute the predicted trajectory of each satellite through the beam pattern of both observations and determine ingress, midpoint, and egress times for the station beam and for  each TAB that the satellite passes through. 

Next, for each satellite passing through the beam pattern, we extracted a time range, centered on the predicted midpoint of the passage, from the dynamic spectra of each of the 91 TABs. The width of this time range was chosen depending on the angular velocity of the satellite on the sky and the FWHM of the station beam, and varied from 12 to 40\,s. Similarly, to increase the signal-to-noise ratio, the extracted dynamic spectra are averaged to a lower time resolution by a factor $n_\mathrm{bin}$ from the native $41.94$\,ms time resolution. We ensured that the duration of the passage through a TAB is covered by at least four averaged time samples.

The extracted dynamic spectra of each TAB were bandpass-calibrated by normalizing with the median of the dynamic spectra of the TABs where the satellite did not pass through. This approach has the advantage that the effect of low-level terrestrial radio frequency interference (RFI), which appears similar in power in all beams, is minimized. After this normalization step, small  variations in intensity ($\sim1\%$) between the different tied-array beams remain, due to differences in sky temperatures and astrophysical sources. We removed these by defining the on-source spectrum as the time range during which the satellite is predicted to be located inside the primary station beam, and the off-source spectrum as the time range it is outside of this area, and further normalizing the spectra by the median intensity of each channel for the off-source time range.

We used the off-source time range to determine the frequency dependent rms of the noise, and used this as input for the radiometer equation to flux-calibrate the normalized dynamic spectra. The radiometer equation relates the rms noise to the spectral flux density of each time and frequency bin in the dynamic spectra through the system temperature and the telescope gain, as well as the time ($t_\mathrm{samp}=41.94\ n_\mathrm{bin}$\,ms) and frequency resolution ($\Delta \nu=12.2$\,kHz) and the number of polarizations recorded ($n_\mathrm{pol}=2$). To determine the frequency dependent system temperature and telescope gain for the LOFAR LBA and HBA observation, we used the method by \citet{kondratiev+2016}, which models the effective area, the station beam model, antenna and sky temperature, together with the beamformer coherency based on the number of stations used to form the tied-array beams, as well as the sky pointing of the beams. This approach is the same as in \citet{divruno+2023}, except for the frequency-dependent calibration, which is strictly necessary for the LBA data.

For each satellite, we used the predicted mid-point times of each beam through which it passed to align the flux-density calibrated dynamic spectra in time, and computed the beam-averaged dynamic spectrum to increase the signal-to-noise ratio and be more sensitive to faint UEMR. Since the satellites do not necessarily pass through the center of each tied-array beam, the average of the flux-density calibrated spectra will underestimate the actual flux density of the satellite compared to a satellite that actually passed through the center of each TAB. Hence, we used the angular response of a TAB (using the FWHM and approximated with a Gaussian) to compute the weighting of all beams and used this to correct the flux density scale. This correction is frequency dependent since the TAB FWHM scales with observing frequency.

The resulting aligned and averaged dynamic spectra were inspected for the presence of both narrowband and broadband emission whose temporal width is consistent with the expected passage duration through the TAB FWHM. In cases where this emission is detected, the temporal profile is fitted with a Gaussian to represent the beam shape, and to provide power flux density measurements and  time offsets due to the satellite running ahead or behind predictions. For narrowband emission, we subtracted the spectral baseline of the surrounding 0.5\,MHz to provide measurements relative to broadband emission. To allow comparison with our earlier measurements from \citet{divruno+2023}, we used the same narrowband frequencies (125, 135, 143.05, 150, and 175\,MHz) and broadband frequency ranges (116--124, 150.05--153, and 157--165\,MHz) for the HBA observation\footnote{The 175\,MHz frequency has become unavailable due to the use of this frequency by digital audio broadcasting.}. For the LBA observation, we chose narrowband frequencies at 25, 50, and 75\,MHz to check for the presence of the harmonics spaced at 25\,MHz intervals that were reported earlier by \citet{divruno+2023}. Broadband frequency ranges correspond to the 37.5--38.25\,MHz and 73--74.6\,MHz frequency ranges where radio astronomy has a secondary and primary ITU-R allocation (see \citetalias{itu_ra769_2} and \citetalias{radioregs}, Vol.\,1, footnote 5.149), respectively. We also include the 50--54\,MHz frequency range, which is assigned for radio amateur usage, and 56--61\,MHz and 61--66\,MHz, where the strongest signals are detected.

The figures in Appendix\,\ref{a:figures} show examples of aligned and averaged dynamic spectra for several satellites in the different observing bands, while the tables in Appendix\,\ref{a:tables} provide lists of the detected Starlink satellites, the properties of the passage through the LOFAR beam pattern, and the power flux density measurements. 

\begin{figure*} 
  \includegraphics[width=\textwidth]{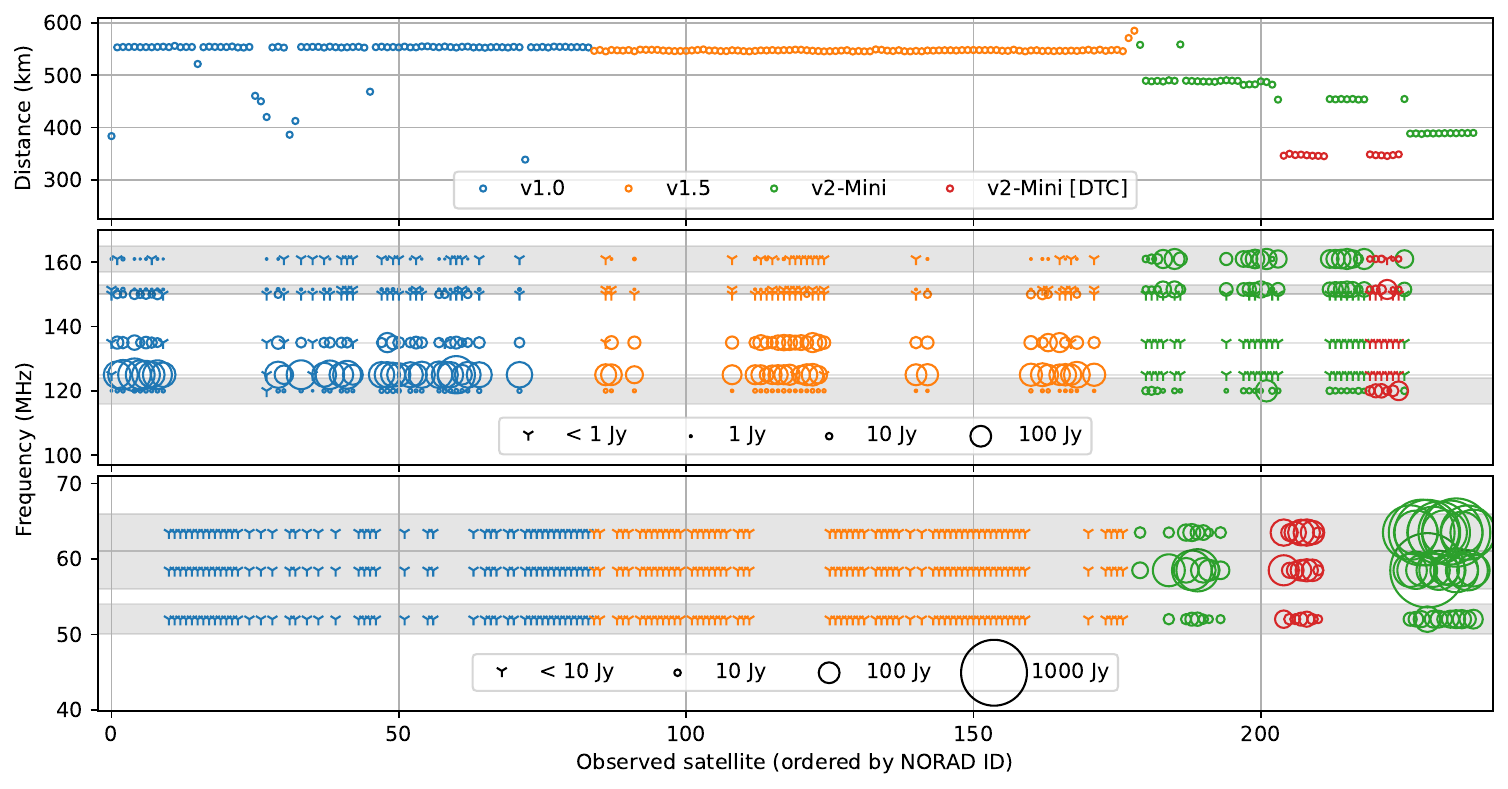}
  \caption{Distances and power flux density measurements of Starlink satellites that passed through the beam pattern of the two 1\,hr LOFAR observations. The horizontal axis denotes the number of satellites that were observed, ordered by their NORAD catalog identifier. As this identifier   sequentially increases with each launched satellite, this axis is essentially ordered in time. All satellites were observed near zenith, and hence their distances are comparable to their orbital altitudes. Power flux density measurements at the different narrowband frequencies or broadband frequency ranges are indicated with circles, where the size of the circle corresponds to the measured flux density. Nondetections are denoted by the $\Ydown$ symbol. The horizontal gray lines and bands indicate the frequency ranges that were used for the flux density determination. There are separate legends for the LBA band from 10--88\,MHz and the HBA band from 110--188\,MHz. The different Starlink satellite versions are indicated with different colors.}
  \label{fig:signal_detections}
\end{figure*}

\section{Results}
A total of 141 Starlink satellites were predicted to pass through at least one tied-array beam in the LBA observation covering 10 to 88\,MHz, while 97 satellites did so for the HBA observation spanning 110 to 188\,MHz. Publicly available information from Jonathan McDowell\footnote{\url{https://planet4589.org/space/con/star/stats.html}} and Gunther Krebs\footnote{\url{https://space.skyrocket.de/doc_sdat/starlink-v1-0.htm} and associated links.} indicates that the observed satellites are of four different versions currently in orbit. Satellite versions v1.0 and v1.5 are from the first generation of Starlink satellites, which were also observed in the 2022 observation presented in \citet{divruno+2023}. The other two satellite versions are from the second generation of Starlink: the normal v2-Mini satellite and the direct-to-cell (DTC) version, which offers cellular mobile phone coverage. Orbital launches of the v2-Mini satellite versions started in February 2023, initially into $43\degr$ inclined orbits, and since August 2023 into orbits with $53\degr$ inclination. Launches with the DTC type v2-Mini's   started in January 2024 and are occasionally launched together with the normal v2-Mini satellites. The DTC version of the v2-Mini satellites are currently all in $53\degr$ inclined orbits.

In the HBA observation, all 97 Starlink satellites are detected, either through narrowband emission, predominantly at 125\,MHz, or broadband emission over parts of or most of the 110--188\,MHz frequency range. The detected signals from the v1.0 and v1.5 satellite versions are consistent in flux density and spectral properties to those of the satellites we detected in the analysis of the 2022 observation in this frequency range \citep{divruno+2023}. The aligned and averaged dynamic spectra of the second-generation v2-Mini satellite versions appear distinctly different, as these do not show the narrowband emission at 125, 135, and 150\,MHz, but instead show significantly brighter broadband UEMR than the v1.0 and v1.5 versions. 

At the LBA frequencies between 10 and 88\,MHz, UEMR from the v2-Mini and v2-Mini DTC versions is clearly detected for 27 of the 29 observed satellites; it is exceedingly bright, reaching power flux densities of hundreds of janskys, and in a few cases even exceeding 1 kJy. The emission is predominantly constrained to a $\sim10$\,MHz band centered around 61\,MHz, but in some cases the broadband emission is detectable down to frequencies of  around 40\,MHz. The broadband UEMR varies in brightness from satellite to satellite, and tends to peak at different frequencies between 60 and 64\,MHz. The v1.0 and v1.5 version satellites are not detected in the LBA data, neither through broadband emission nor at the narrowband frequencies at 25, 50, and 75\,MHz. In the few cases where signals were present in the aligned and averaged spectra, they could be explained by the strong UEMR from v2-Mini or v2-Mini DTC satellites that were also in or near the station beam at that time as a result of the increasing density of satellites in the sky. As the v1.0 and v1.5 satellite versions emit narrowband UEMR at 125, 150, and 175\,MHz, we suggested in \citet{divruno+2023} that these may be harmonics of a 25\,MHz clock signal on board the satellite, and that we would expect emission at the fundamental frequency of 25\,MHz and the harmonics at 50 and 75\,MHz. While the 25\,MHz frequency is lost due to terrestrial RFI, the LBA observations show that if present, the narrow- or broadband UEMR at 50 and 75\,MHz must have power flux densities below $S_\nu < 10$\,Jy ($3\sigma$). 

\begin{figure} 
  \includegraphics[width=\columnwidth]{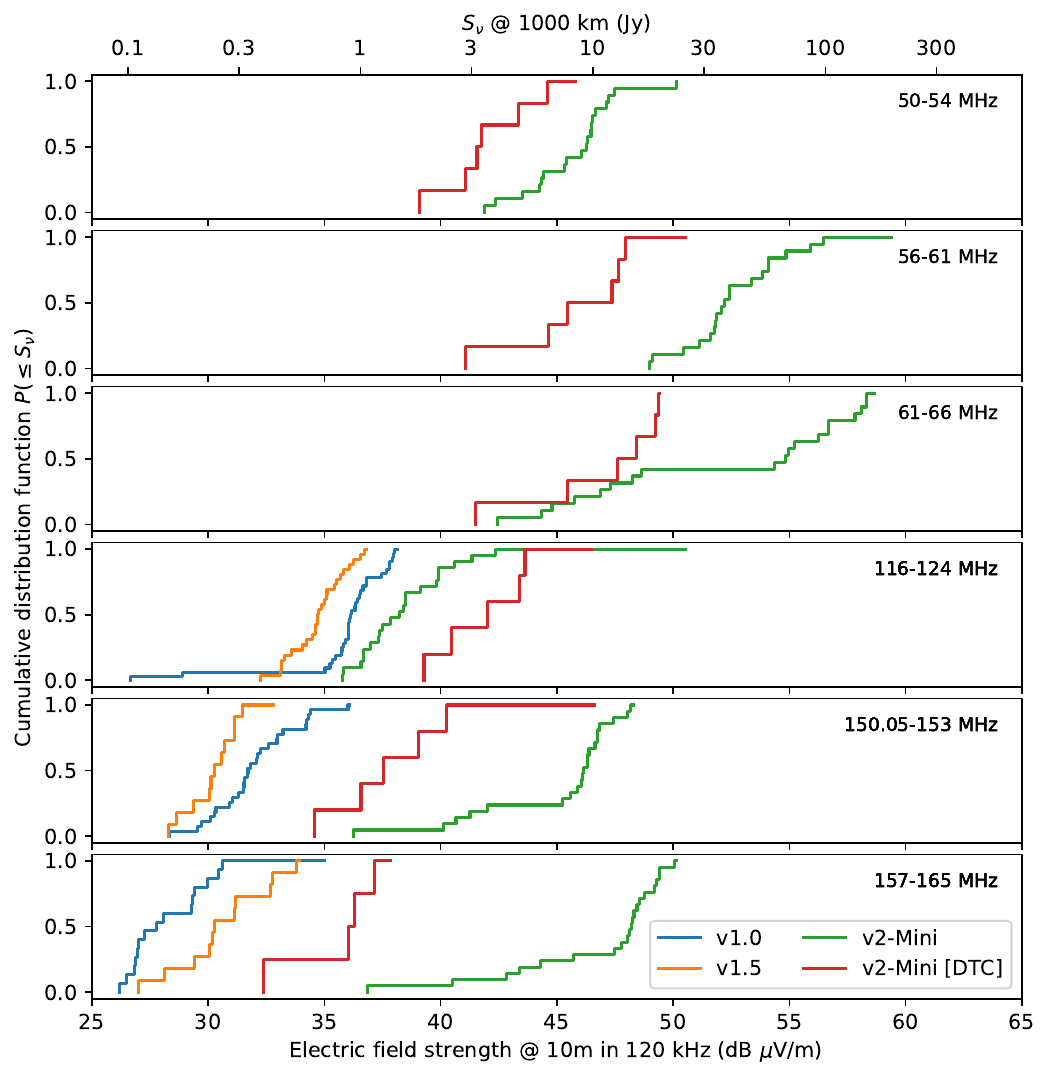}
  \caption{Comparison of the intrinsic UEMR power flux densities for different broadband frequency ranges per Starlink satellite version. Each panel shows a cumulative distribution function of the power flux density measurements scaled to a distance of 1000\,km (top axis) or represented as the electric field strength measured by a detector at 10\,m distance using 120\,kHz of bandwidth (bottom axis).}
  \label{fig:efields}
\end{figure}

Figure\,\ref{fig:signal_detections} provides an overview of the power flux density measurements at the different narrowband frequencies and broadband frequency ranges, their distances to the telescope at the time of detection, and the satellite versions. It is clear that the power flux densities of the v2-Mini and v2-Mini DTC Starlink satellites are higher than those of the first generation. However, the v2-Mini and v2-Mini DTC versions were observed at smaller distances as these satellites operate at lower orbital altitudes. To determine whether the UEMR emitted by the second generation of Starlink satellites is intrinsically brighter, we corrected the observed power flux densities by scaling them to a fixed distance of 1000\,km, as shown in Fig.\,\ref{fig:efields}. This approach also has the advantage that these normalized power flux densities can be directly related to the electric field strength emitted by the satellites, if received by a detector at a distance of 10\,m and integrating over a 120\,kHz bandwidth. This implicitly assumes that the emitted UEMR is isotropic, which likely is not the case, but allows   further comparison to commercial electromagnetic compatibility (EMC) standards that we used in \citet{divruno+2023}. From Fig.\,\ref{fig:efields} we find that intrinsic levels of broadband UEMR emitted by the observed satellites from the second generation of Starlink (the v2-Mini and v2-Mini DTC versions) are higher than those observed from the  first-generation satellites. 

In the LBA band, UEMR from the observed v2-Mini and v2-Mini DTC satellites shows spectral structure over the entire frequency range where UEMR is detected. This structure consists of a ``comb'' of regularly spaced peaks in frequency. Power spectra of this emission at the 12.2\,kHz frequency resolution between 56 and 66\,MHz show significant peaks at multiple, harmonically related, peaks at frequencies of 27.5, 36.66, 55, 110, and 220\,kHz for the v2-Mini satellites, and at 37.5, 50, 75, and 150\,kHz for the v2-Mini DTC satellites. This shows that this spacing is distinctly different between the v2-Mini and v2-Mini DTC satellite versions that were observed. Due to variations in the power of the spectral harmonics, we cannot identify a fundamental frequency of these combs. Spectral structure is less apparent for the UEMR detections in the HBA band. Power spectra over the two frequency ranges between 116--124 and 157--165\,MHz show that some of the version v1.5 satellites observed have a comb with a fundamental frequency at 50\,kHz in the 157--165\,MHz band, similar to what was seen in our earlier observations of satellites of this version \citep{divruno+2023}. The v2-Mini satellites observed predominantly show periodic signals at frequencies of 48.8, 65, and 97.5\,kHz in the 157 to 165\,MHz band, while most of the v2-Mini DTC satellites show  a comb with 50 or 150\,kHz spacing in the lower band from 116 to 124\,MHz. For these combs, no fundamental frequency can be identified due to the satellite-to-satellite power variations of the harmonics.

\section{Discussion and conclusions}
We find that the second generation of Starlink satellites that we observed with LOFAR emit higher levels of unintended electromagnetic radiation (UEMR) over a broader frequency range compared to that emitted by the first generation of Starlink satellites. Our observations show that in the 150.05--153\,MHz primary radio astronomy band, the broadband UEMR of the second-generation v2-Mini and v2-Mini DTC satellites is, on average, 15\,dB and 7\,dB brighter than that of the first-generation v1.0 and v1.5 Starlink satellites. On a linear scale, this corresponds to factors of 32 and 5, respectively. As is evident from Fig.\,\ref{fig:efields}, this trend is also present in the 116 to 124\,MHz and 157 to 165\,MHz bands, and in the frequency bands from 50 to 66\,MHz where the satellites from the first generation are not detected. On the other hand, the strong narrowband UEMR that is seen in the v1.0 and v1.5 satellites at 125, 135, and 150\,MHz appears to be absent in the v2-Mini and v2-Mini DTC satellites. While this is an improvement, it is completely negated by the stronger broadband UEMR, which affects a significantly larger part of the observed frequency range.

The issue of the higher levels of UEMR from the second-generation Starlink satellites is further exacerbated by the lower orbits in which these satellites operate. These satellites are used in the (modified) Generation 2 Starlink constellation, for which the US Federal Communications Commission (FCC) has approved operational orbits at 448 and 482\,km for the v2-Mini satellites, and 360\,km for the v2-Mini DTC satellites. As a result of these lower orbits and resulting smaller distances to Earth-based telescopes, the signals will be 30 to 130\% brighter compared to the Generation 1 Starlink constellation, which mostly operates at orbital altitudes of 550\,km.

In \citet{divruno+2023}, we used the ITU-R recommended equivalent power flux density (EPFD) method \citepalias{itu_m1583_1,itu_s1586_1} to simulate the aggregate impact of large numbers of satellites in several satellite constellations and estimate their compatibility with ITU-R recommended interference thresholds. For the Generation 1 Starlink constellation of 4408 satellites in orbits at 550\,km altitude, we found that the intrinsic electric field strength of an individual satellite had to remain below $11.7$\,dB [$\upmu$V\,m$^{-1}$] to satisfy the $-194$\,dB [W\,m$^{-2}$] ITU-R threshold for the radio astronomy band of 150.05--153\,MHz \citepalias{itu_ra769_2}. The intrinsic broadband UEMR from observed Starlink v1.0 and v1.5 satellites from the 2022 observation had electric field strengths  of 21 to 39 dB [$\upmu$V\,m$^{-1}$], already significantly exceeding that limit. These values also exceed typical commercial EMC standards (e.g.,\ 30 dB [$\upmu$V\,m$^{-1}$] from CISPR, see discussion in \citealt{divruno+2023}). Given that the Generation 2 Starlink constellation will consist of even more satellites than the Generation 1 constellation, that these satellites will be operating at lower orbital altitudes, and that this constellation will consist of the v2-Mini and v2-Mini DTC satellites that are now found to emit even stronger UEMR, we can conclude that the \citetalias{itu_ra769_2} recommended interference threshold levels are exceeded even further in this radio astronomy band. The LOFAR observations presented here do not detect UEMR in the 73--74.6\,MHz band allocated to radio astronomy. However, preliminary analysis of imaging observations of second-generation Starlink satellites with the NenuFAR telescope in France \citep{zarka+2012}, indicates that some may be detectable in that band (Zhang et al., in prep.). While they are beyond the scope of this paper, EPFD simulations of the Generation 2 Starlink constellation would be required to estimate the intrinsic electric field strengths required to keep the aggregate emission of UEMR of this constellation in accordance with the ITU-R recommendations. 

Current wording of the ITU-R radio regulations \citepalias{radioregs} and recommendations \citepalias[e.g.,][and related recommendations]{itu_ra769_2} discuss emissions in terms of wanted and unwanted emission related to signal transmission, where the unwanted emission is a byproduct of the wanted emission, for example due to  out-of-band emission in the spectral domain. The UEMR as defined in our earlier paper \citep{divruno+2023} appears to fall outside of these regulations. As such, UEMR is not subject to the ITU-R interference limits which protect certain parts of the spectrum for radio astronomical applications. Hence, we reiterate our earlier recommendation that UEMR from satellites should be considered in the regulatory processes.

The impact of the observed UEMR on radio astronomy likely varies between different science cases. The first-order effect will be a loss of sensitivity of low-frequency radio telescopes since the time and frequency ranges within an observation that are affected by satellite UEMR may have to be preemptively masked. However, given that low-frequency radio telescopes are primarily built for their large fields of view, the large numbers of satellites from current and future satellite constellations may lead to the situation that one or more satellites are present in the telescope's field of view at any given time. In this case, temporal masking of data will no longer provide useful data. This is the primary reason why broadband UEMR is particularly worrisome for radio astronomy; it increases the risk that the entire observing bandwidth is affected by UEMR for the entire duration of the observation. A second-order effect, primarily affecting interferometric telescope arrays, is that for closely spaced array elements (parabolic dishes or antenna stations), satellites will appear at the same sky location. As a result, UEMR will not decorrelate on the shortest baselines between individual array elements and may introduce artifacts on large spatial scales.

In contrast with this situation, astronomical radio observatories put a great deal of effort in mitigating their internally generated UEMR in all frequencies covered by their telescopes, as electrical devices needed to run telescopes are also prone to producing radio noise. Observatories such as LOFAR and the SKA Observatory go to great lengths, imposing extremely tight radio emission requirements on each  of the subsystems that comprise the telescopes. In the SKA-Low case in Western Australia, some examples of this are the Central Processing Facility building, which is designed to shield all the computing equipment that performs the first processing of SKA-Low data,  and the power and signal distribution boxes located in close proximity to SKA-Low antennas, having requirements that are more than 100\,dB ($10^{10}$ times) stricter than commercial standards for radiated emissions.

The UEMR from equipment close to, but not associated with radio telescopes is a day-to-day reality. Like the UEMR from satellites, their spectra are generally tens to hundreds of MHz wide, and have a comb-like structure in addition to a more diffuse wide-band spectrum. This type of UEMR is typically handled in various ways, for example   i) raising awareness with local stakeholders; ii) advocating for local, regional, and/or national protected geographic radio quiet zones; iii) establishing bilateral covenants requiring EMC limits stricter than typical levels specified in industry norms such as \citetalias{cispr32} and \citetalias{en55032} (e.g.,\ wind and solar photovoltaic installations near the LOFAR core\footnote{\url{https://www.rvo.nl/onderwerpen/bureau-energieprojecten/lopende-projecten/windpark-dm-en-om}}); iv) cooperation with equipment owners to mitigate at the source (mending electric fences, replacing LED lights, switching off CCTV cameras); and/or v) formal regulatory complaints and follow-up by national administrations in cases of noncooperative parties. The last step is generally only viable if the sources exceed EMC norms similar to \citetalias{cispr32}/\citetalias{en55032}. As far as we are aware, a similar regulatory or normative framework is lacking for space applications. In contrast to satellites, we note that all of these sources are generally greatly attenuated by the telescope as, unlike satellites, they are never directly in its main beam.

In the absence of regulations that address UEMR emission from satellites, the astronomical community will have to  raise and address this issue with regulatory bodies as well as satellite operators, and must continue to do so. Fortunately, SpaceX/Starlink is already actively co-operating with both optical astronomy \citep[e.g.,][]{tyson+2020} and radio astronomy \citep[e.g.,][]{nhan+2024} to investigate and/or test mitigation strategies. 
Our observations and analysis, presenting properties of the UEMR (electric field strengths, emission frequencies, comb properties) of different satellite versions, possibly combined with those of other radio telescopes, may provide information that allows SpaceX/Starlink to identify the satellite components involved in the emission of UEMR and devise mitigation strategies in already operational satellites, as well as future designs of the hardware. 

\begin{acknowledgements}
We thank Colin Lonsdale for informing us of the presence of UEMR from the second-generation Starlink satellites. We acknowledge fruitful discussions with Jess Dempsey, Michiel van Haarlem and Wim van Cappellen.

This paper is based on data obtained with the International LOFAR Telescope (ILT) under project code LC20\_009. LOFAR \citep{vanhaarlem+2013} is the Low Frequency Array designed and constructed by ASTRON. It has observing, data processing, and data storage facilities in several countries, that are owned by various parties (each with their own funding sources), and that are collectively operated by the ILT foundation under a joint scientific policy. The ILT resources have benefitted from the following recent major funding sources: CNRS-INSU, Observatoire de Paris and Université d'Orléans, France; BMBF, MIWF-NRW, MPG, Germany; Science Foundation Ireland (SFI), Department of Business, Enterprise and Innovation (DBEI), Ireland; NWO, The Netherlands; The Science and Technology Facilities Council, UK; Ministry of Science and Higher Education, Poland. The project leading to this publication has received funding from the European Union’s Horizon 2020 research and innovation programme under grant agreement No 101004719.

This paper made extensive use of the Python scientific software stack, and we acknowledge the developers of \texttt{numpy} \citep{NumPy}, \texttt{matplotlib} \citep{Matplotlib}, \texttt{scipy} \citep{SciPy}, \texttt{astropy} \citep{Astropy2013,Astropy2022} and \texttt{Skyfield} \citep{rhodes2019}. 

\end{acknowledgements}


\onecolumn
\begin{appendix}
\section{Additional figures}\label{a:figures}
\begin{figure*}[!h]
  \includegraphics[width=\textwidth]{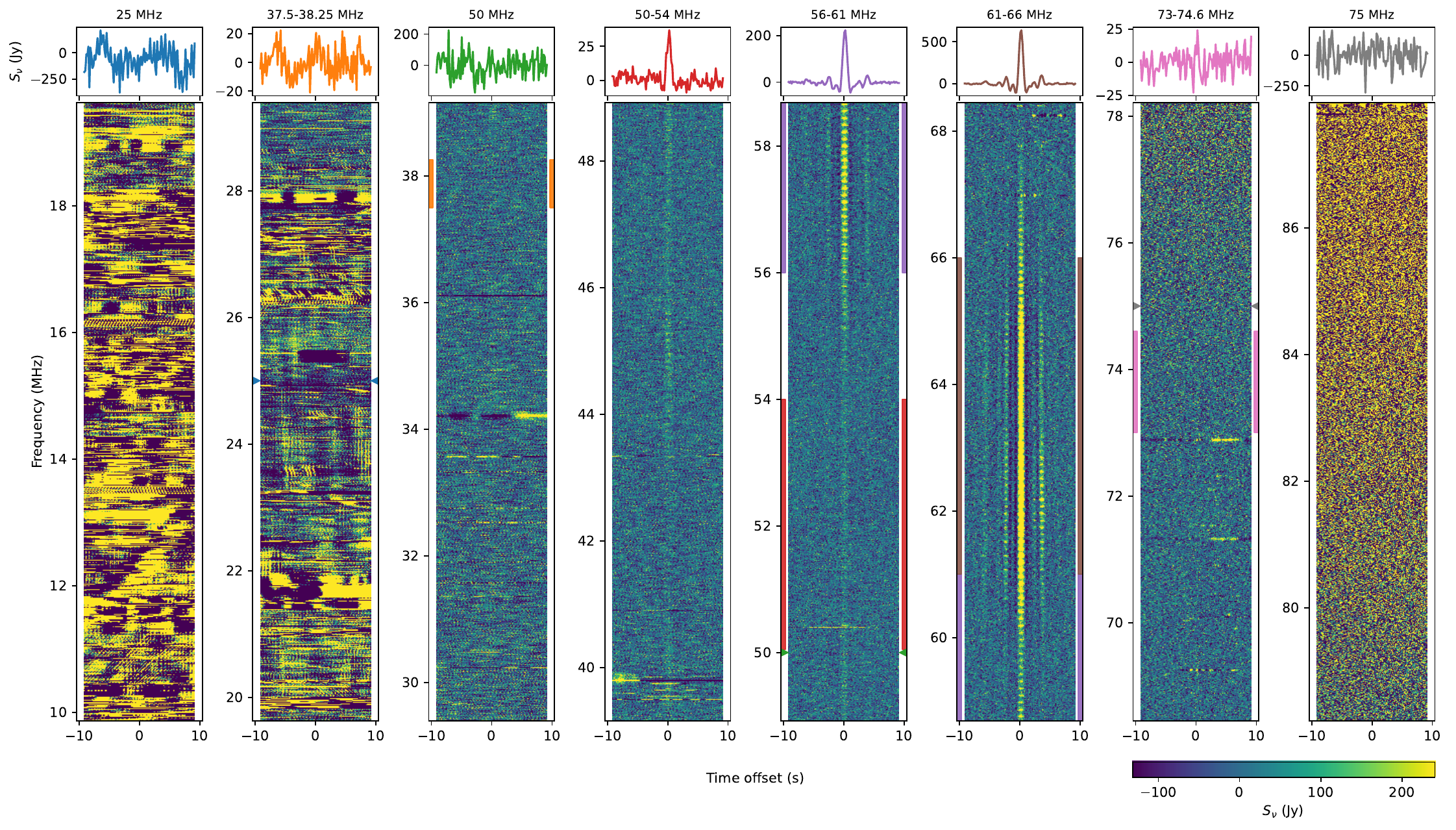}
  \caption{Spectral and temporal properties of the passage of the Starlink v2-Mini satellite Starlink-31441 [60091/2024-117A] (average of 11 TABs) in the LBA band from 10 to 88\,MHz. Normalized, aligned, and averaged dynamic spectra (in power flux density units) are shown over the entire observed bandwidth and are centered on the predicted passage time of the satellite. Time series at specific narrowband frequencies and broadband frequency ranges are shown in the top insets. The color of each time series matches the marked frequencies and frequency ranges, in the same colors as the sides of the dynamic spectra.}
  \label{fig:zoomed_plot_lba1}
\end{figure*}

\begin{figure*}[!h]
  \includegraphics[width=\textwidth]{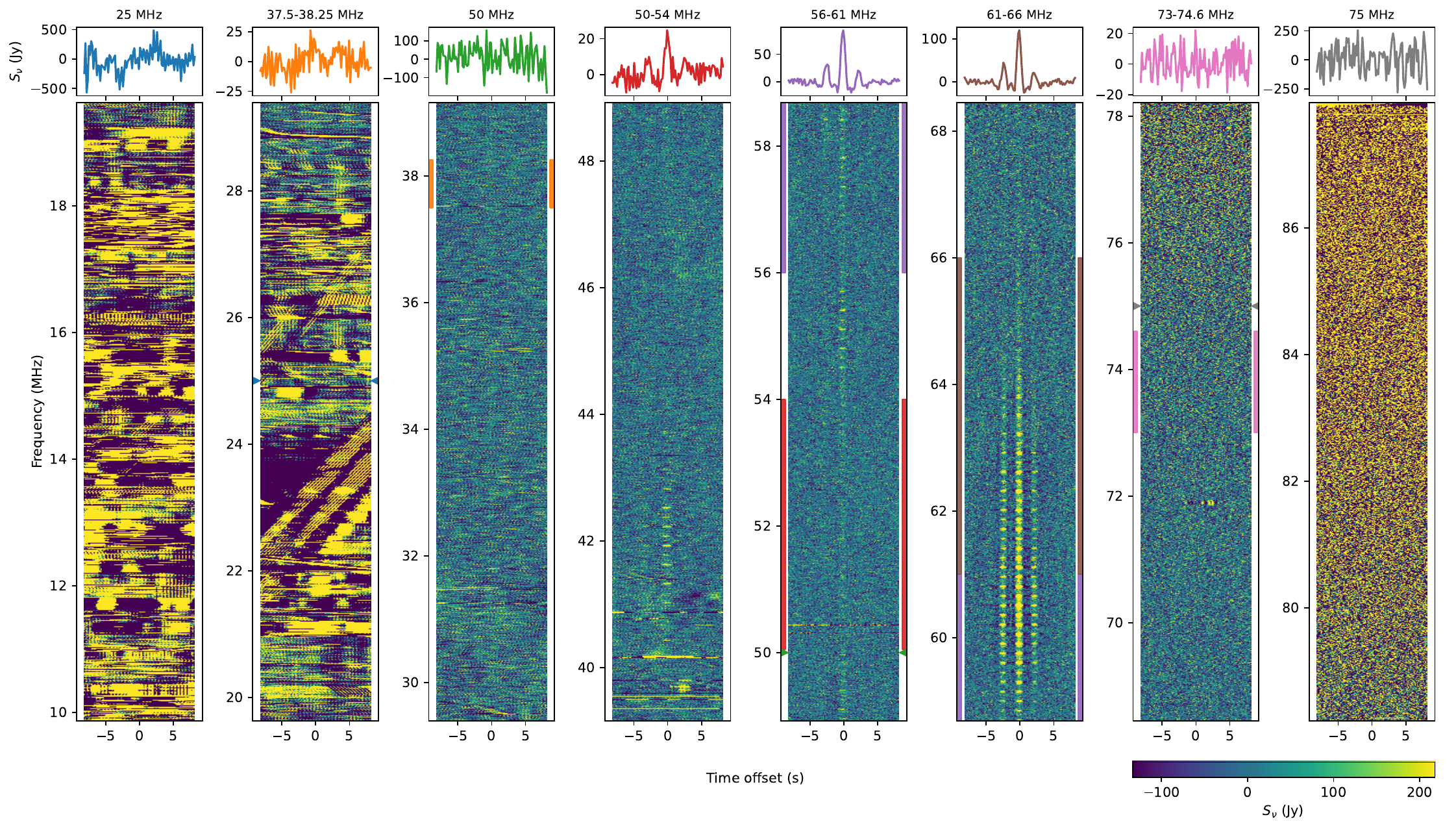}
  \caption{Spectral and temporal properties of the passage of the Starlink v2-Mini DTC satellite Starlink-11133 [DTC] [59954/2024-107K] (average of 11 TABs) in the LBA band from 10 to 88\,MHz.}
  \label{fig:zoomed_plot_lba2}
\end{figure*}

\begin{figure*}[!h]
  \includegraphics[width=\textwidth]{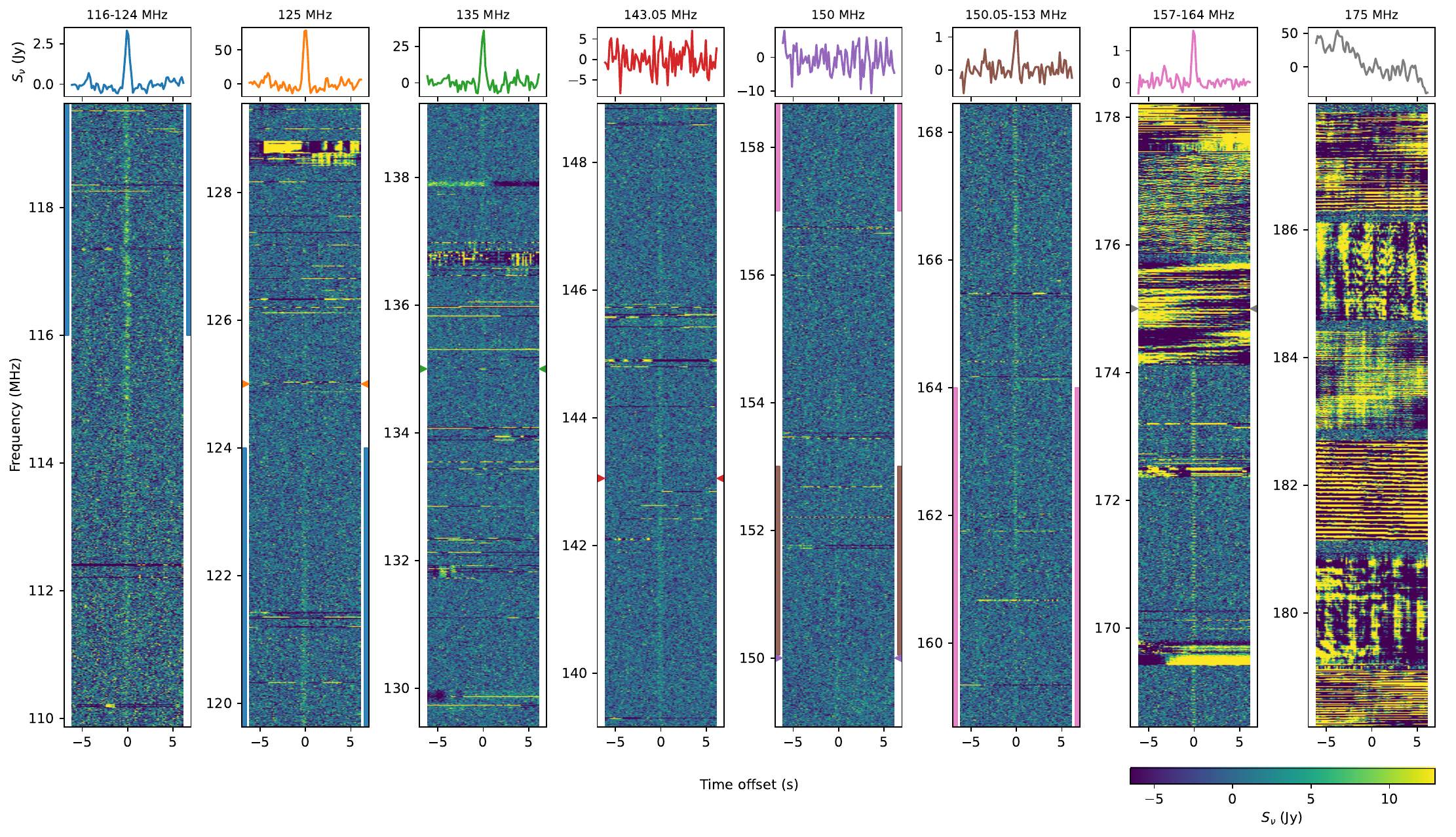}
  \caption{Spectral and temporal properties of the passage of of Starlink v1.5 satellite Starlink-3349 [50813/2022-001L] (average of 11 TABs) in the HBA band from 110 to 188\,MHz.}
  \label{fig:zoomed_plot_hba1}
\end{figure*}

\begin{figure*}[!h]
  \includegraphics[width=\textwidth]{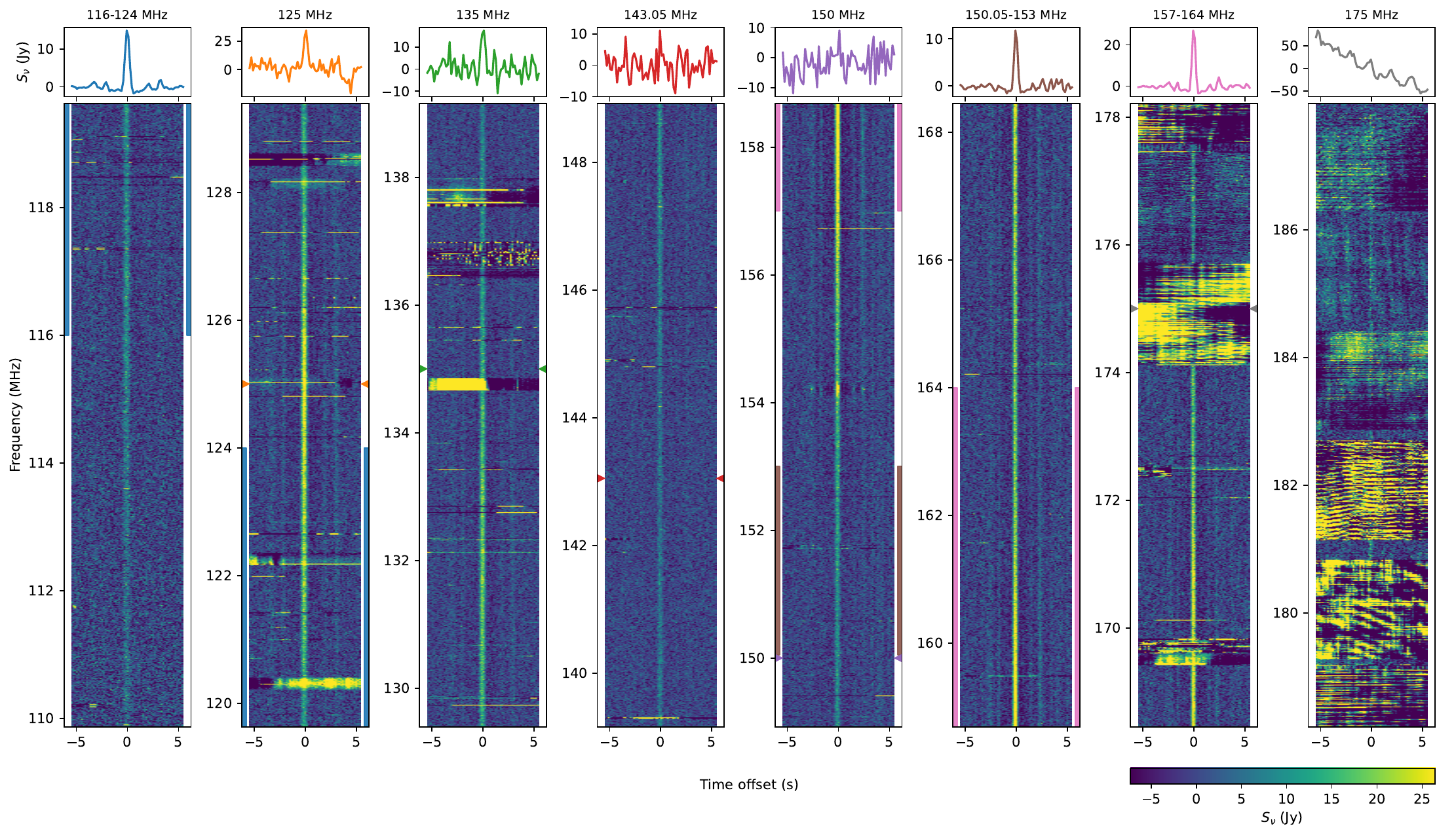}
  \caption{Spectral and temporal properties of the passage of of Starlink v2-Mini satellite Starlink-30518 [58029/2023-156B] (average of 11 TABs) in the HBA band from 110 to 188\,MHz.}
  \label{fig:zoomed_plot_hba2}
\end{figure*}

This appendix shows figures of aligned and averaged dynamic spectra for Starlink satellites in which UEMR is detected. Figures\,\ref{fig:zoomed_plot_lba1} and \,\ref{fig:zoomed_plot_lba2} show detections for Starlink v2-Mini and v2-Mini DTC satellites in the LBA observing band between 10 and 88\,MHz.
In the HBA band covering 110 to 188\,MHz, Figs.\,\ref{fig:zoomed_plot_hba1} and \ref{fig:zoomed_plot_hba2} show aligned and averaged dynamic spectra for a Starlink v1.5 and a v2-Mini satellite. The color scale of the dynamic spectra is in power flux density.

These dynamic spectra show the UEMR from Starlink satellites, as well as interference from terrestrial sources. Satellite UEMR will stand out because of its temporal signature as the satellite passes through the telescope field-of-view. For comparison, terrestrial RFI is primarily detected through the telescope side-lobes and hence has no specific, nor predictable, variation with time. 

Terrestrial RFI in the LOFAR band is primarily due to allocated services, and an overview of these is given in \citet{offringa+2013}. In the LBA band from 10 to 88\,MHz, these services predominantly affect frequencies below 30\,MHz, where beyond the horizon transmissions are detectable through reflections off the ionosphere. In the HBA band from 110 to 188\,MHz, digital audio broadcasting channels continuously occupy frequencies in several 1.6\,MHz wide bands from 174\,MHz and higher, while air traffic control (118 to 137\,MHz), satellite downlinks (137 to 138\,MHz) and amateur radio (144 to 146\,MHz) use smaller bandwidths (a few kHz) and transmit for limited time periods (seconds to minutes).

\newpage
\section{Tables}\label{a:tables}
\begin{table*}[!h]
\tiny
\centering
\caption{Second-generation Starlink satellites that have been detected between 10 and 88\,MHz. }
\label{tab:satellite_passes_lba}
\begin{tabular}{ll|rrr|ccc}
\hline\hline
NORAD / COSPAR & Name & $d$ & $t_\mathrm{mid}$ & $n_\mathrm{TAB}$ & \multicolumn{3}{c}{$S_\nu$ (Jy)} \\
& &  (km) & (s) & & 50--54 & 56--61 & 61--66  \\
& &       &     & & (MHz) & (MHz) & (MHz) \\
\hline
57946/2023-148M & STARLINK-30704 & 558.0 & 740.14 & 11 &      & 57.9 & 26.8 \\
58532/2023-192B & STARLINK-31035 & 490.1 & 3322.12 & 11 & 25.5 & 235.8 & 25.1 \\
58543/2023-192N & STARLINK-31049 & 489.4 & 3291.96 & 11 & 24.6 & 142.0 & 61.7 \\
58544/2023-192P & STARLINK-31048 & 488.9 & 2902.19 & 11 & 39.4 & 361.8 & 67.6 \\
58545/2023-192Q & STARLINK-31057 & 488.3 & 2875.42 & 11 & 41.1 & 413.0 & 45.1 \\
58546/2023-192R & STARLINK-31050 & 487.8 & 2455.88 & 10 & 20.9 & 140.9 & 50.0 \\
58548/2023-192T & STARLINK-31016 & 487.6 & 2063.28 & 10 & 14.4 & 102.7 & 16.4 \\
58549/2023-192U & STARLINK-31053 & 487.4 & 2036.87 & 2 &      &      &      \\
58551/2023-192W & STARLINK-30999 & 489.3 & 1954.86 & 8 & 15.9 & 73.0 & 27.8 \\
59949/2024-107E & STARLINK-11138 [DTC] & 345.9 & 2762.60 & 11 & 70.3 & 209.1 & 160.4 \\
59950/2024-107F & STARLINK-11130 [DTC] & 349.5 & 3290.90 & 10 & 27.1 & 52.7 & 63.5 \\
59951/2024-107G & STARLINK-11126 [DTC] & 347.2 & 3110.70 & 11 & 23.5 & 64.7 & 106.2 \\
59952/2024-107H & STARLINK-11124 [DTC] & 347.9 & 2928.68 & 11 & 39.5 & 99.7 & 153.5 \\
59953/2024-107J & STARLINK-11143 [DTC] & 346.8 & 2749.49 & 11 & 53.1 & 115.0 & 160.7 \\
59954/2024-107K & STARLINK-11133 [DTC] & 345.9 & 2569.45 & 11 & 26.5 & 107.9 & 128.5 \\
59955/2024-107L & STARLINK-11144 [DTC] & 345.6 & 2389.55 & 10 & 15.0 & 23.8 & 26.2 \\
59956/2024-107M & STARLINK-11134 [DTC] & 345.0 & 2209.46 & 2 &      &      &      \\
60091/2024-117A & STARLINK-31441 & 388.2 & 353.10 & 11 & 40.0 & 255.8 & 686.8 \\
60092/2024-117B & STARLINK-32044 & 388.5 & 322.46 & 11 & 50.2 & 319.8 & 445.8 \\
60093/2024-117C & STARLINK-31239 & 387.6 & 279.19 & 11 & 59.4 & 221.7 & 996.7 \\
60094/2024-117D & STARLINK-32007 & 388.7 & 261.57 & 11 & 151.5 & 1273.6 & 990.3 \\
60095/2024-117E & STARLINK-32028 & 388.5 & 231.09 & 11 & 65.2 & 255.5 & 620.5 \\
60096/2024-117F & STARLINK-32016 & 388.6 & 200.64 & 11 & 65.9 & 354.8 & 887.4 \\
60097/2024-117G & STARLINK-32003 & 388.9 & 170.23 & 11 & 50.7 & 189.2 & 458.5 \\
60098/2024-117H & STARLINK-32026 & 389.0 & 109.19 & 11 & 67.9 & 210.9 & 941.7 \\
60099/2024-117J & STARLINK-31732 & 388.9 & 139.72 & 11 & 75.5 & 447.3 & 1074.8 \\
60100/2024-117K & STARLINK-31996 & 389.3 & 78.75 & 11 & 77.3 & 235.1 & 399.1 \\
60101/2024-117L & STARLINK-31999 & 389.2 & 48.16 & 11 & 62.5 & 375.3 & 680.8 \\
60102/2024-117M & STARLINK-31954 & 389.6 & 12.58 & 11 & 81.8 & 242.4 & 486.1 \\
\hline
\end{tabular}
\tablefoot{Satellites are identified by their NORAD catalog numer and the COSPAR international designator, which provides the launch year, launch number, and sequential identifier, and by their name. Satellites with [DTC] in their name are of the v2-Mini direct-to-cell (DTC) version; all other satellites are of the v2-Mini version. For each satellite the distance ($d$ in km), passage mid-point ($t_\mathrm{mid}$ in s) during the 1\,hr observation (which started on July 19, 2024, at 06:00\,UTC) and the number of tied-array beams the satellite passed through ($n_\mathrm{TAB}$) are provided. Power flux density measurements ($S_\nu$ in Jy) are given for three frequency ranges.}
\end{table*}

\begin{table*}[!h]
\tiny
\centering
\caption{First-generation Starlink satellites that have been detected between 110 and 188\,MHz. }
\label{tab:satellite_passes_hba_gen1}
\begin{tabular}{ll|rrr|cccccc}
\hline\hline
NORAD / COSPAR & Name & $d$ & $t_\mathrm{mid}$ & $n_\mathrm{TAB}$ & \multicolumn{6}{c}{$S_\nu$ (Jy)} \\
& &  (km) & (s) & & 116--124 & 125 & 135 & 150 & 150.05--153 & 157--165 \\
& &       &     & & (MHz) & (MHz) & (MHz) & (MHz) & (MHz) & (MHz) \\
\hline
\textit{Starlink v1.0} & & & & & & & & & & \\
45103/2020-006BM & STARLINK-1196 & 383.4 & 539.83 & 11 & 0.7 &      &      &      &      & 0.9 \\
45362/2020-019C & STARLINK-1306 & 553.2 & 2469.55 & 11 & 2.9 & 167.3 & 34.7 & 14.9 & 0.9 &      \\
45366/2020-019G & STARLINK-1262 & 553.9 & 1146.02 & 11 & 3.0 & 200.7 & 30.8 & 10.1 & 2.9 & 2.3 \\
45367/2020-019H & STARLINK-1273 & 553.8 & 3131.25 & 2 & 2.9 &      &      &      & 1.3 &      \\
45368/2020-019J & STARLINK-1276 & 554.2 & 485.78 & 11 & 3.3 & 244.4 & 48.7 & 22.1 & 1.0 & 0.6 \\
45373/2020-019P & STARLINK-1295 & 553.6 & 1808.10 & 11 & 2.9 & 191.6 & 17.0 & 12.8 & 0.5 & 0.4 \\
45374/2020-019Q & STARLINK-1300 & 553.7 & 1477.32 & 11 & 3.2 & 178.4 & 31.6 & 20.0 & 1.3 & 0.7 \\
45375/2020-019R & STARLINK-1302 & 553.5 & 2800.60 & 11 & 2.5 & 153.9 & 25.2 & 10.5 & 0.8 &      \\
45377/2020-019T & STARLINK-1305 & 554.0 & 815.43 & 11 & 2.9 & 192.2 & 18.9 & 23.8 & 1.2 & 0.8 \\
45379/2020-019V & STARLINK-1319 & 554.5 & 155.49 & 11 & 2.7 & 139.2 &      &      & 0.9 & 0.4 \\
45770/2020-038AS & STARLINK-1462 & 419.9 & 3104.90 & 11 &      &      &      &      & 1.3 & 0.7 \\
46542/2020-070L & STARLINK-1692 & 554.2 & 3280.05 & 12 & 3.1 & 159.1 & 35.9 & 12.3 & 2.0 & 0.6 \\
46544/2020-070N & STARLINK-1694 & 552.8 & 1733.87 & 6 & 2.5 & 79.2 &      &      & 0.7 &      \\
46550/2020-070U & STARLINK-1701 & 554.0 & 2950.69 & 11 & 2.7 & 192.3 & 23.0 &      & 1.1 &      \\
46554/2020-070Y & STARLINK-1671 & 554.1 & 743.25 & 11 & 0.6 &      &      &      & 0.7 &      \\
46561/2020-070AF & STARLINK-1709 & 552.9 & 1648.75 & 9 & 2.8 & 134.4 & 21.7 &      & 1.2 &      \\
46563/2020-070AH & STARLINK-1730 & 554.5 & 4.94 & 11 & 3.1 & 187.0 & 24.6 &      & 1.1 & 0.3 \\
46572/2020-070AS & STARLINK-1531 & 552.9 & 2204.51 & 6 & 2.9 & 131.1 & 24.2 &      &      &      \\
46577/2020-070AX & STARLINK-1682 & 553.3 & 1071.23 & 11 & 3.3 & 178.7 & 23.1 &      &      &      \\
46586/2020-070BG & STARLINK-1737 & 553.8 & 335.48 & 9 & 2.4 & 91.9 &      &      &      &      \\
48093/2021-027B & STARLINK-2404 & 554.5 & 410.49 & 11 & 4.5 & 150.2 & 21.2 &      & 1.5 &      \\
48094/2021-027C & STARLINK-2412 & 553.4 & 1883.94 & 11 & 4.6 & 149.2 & 86.2 &      & 1.4 & 0.8 \\
48095/2021-027D & STARLINK-2414 & 553.9 & 2543.80 & 11 & 4.7 & 135.1 & 40.6 &      & 1.9 &      \\
48099/2021-027H & STARLINK-2428 & 553.2 & 1222.05 & 11 & 4.2 & 125.0 & 29.7 &      & 1.1 &      \\
48111/2021-027V & STARLINK-2448 & 553.2 & 1402.33 & 11 & 4.0 & 146.0 & 22.2 &      & 1.4 & 0.3 \\
48113/2021-027X & STARLINK-2450 & 553.3 & 891.38 & 3 & 2.3 & 107.7 & 32.5 &      &      &      \\
48115/2021-027Z & STARLINK-2452 & 555.0 & 79.21 & 11 & 3.5 & 158.9 & 26.2 &      & 1.2 & 0.5 \\
48130/2021-027AQ & STARLINK-2472 & 553.4 & 2620.26 & 10 & 2.8 & 163.7 & 21.4 & 13.9 & 1.0 & 0.6 \\
48133/2021-027AT & STARLINK-2475 & 554.7 & 3536.29 & 11 & 4.3 & 158.2 &      & 13.3 & 1.0 & 0.3 \\
48134/2021-027AU & STARLINK-2476 & 553.7 & 2213.20 & 11 & 3.2 & 148.7 & 28.5 &      & 0.8 &      \\
48135/2021-027AV & STARLINK-2478 & 552.9 & 2288.87 & 6 & 3.5 & 321.6 & 33.3 &      & 2.9 &      \\
48136/2021-027AW & STARLINK-2479 & 554.3 & 2875.03 & 11 & 2.9 & 89.8 & 16.8 &      &      &      \\
48138/2021-027AY & STARLINK-2481 & 554.6 & 3205.66 & 11 & 2.9 & 154.1 & 18.6 & 14.0 & 1.0 & 0.3 \\
48141/2021-027BB & STARLINK-2484 & 553.3 & 1552.86 & 11 & 4.5 & 145.2 & 28.8 &      & 2.0 &      \\
48435/2021-040H & STARLINK-2633 & 553.4 & 2139.10 & 11 & 4.3 & 150.6 & 22.5 &      & 1.9 &      \\[0.5em]
\textit{Starlink v1.5} & & & & & & & & & & \\
49730/2021-115G & STARLINK-3249 & 545.5 & 1874.18 & 3 & 3.6 & 102.2 &      &      &      &      \\
49733/2021-115K & STARLINK-3252 & 548.2 & 553.31 & 7 & 2.2 & 100.2 & 38.3 &      &      & 0.8 \\
49739/2021-115R & STARLINK-3155 & 545.8 & 1543.50 & 11 & 2.4 & 67.4 & 34.1 &      & 0.8 & 1.8 \\
50803/2022-001A & STARLINK-3321 & 547.6 & 1032.03 & 3 & 1.6 & 83.6 & 34.5 &      &      &      \\
50813/2022-001L & STARLINK-3349 & 546.4 & 1213.78 & 11 & 3.5 & 83.6 & 30.9 &      & 1.0 & 1.4 \\
50815/2022-001N & STARLINK-3234 & 547.4 & 701.90 & 9 & 2.3 & 83.5 & 51.5 &      & 0.6 &      \\
50817/2022-001Q & STARLINK-3341 & 547.3 & 371.80 & 11 & 3.0 & 43.4 & 33.7 &      & 1.0 & 1.8 \\
50822/2022-001V & STARLINK-3343 & 548.0 & 3014.22 & 4 & 2.4 & 80.0 & 30.4 &      &      &      \\
50823/2022-001W & STARLINK-3334 & 547.3 & 40.97 & 11 & 2.0 & 86.1 & 48.2 &      & 0.8 & 0.4 \\
50825/2022-001Y & STARLINK-3332 & 547.9 & 3344.32 & 11 & 3.1 & 84.5 & 55.2 &      &      & 1.4 \\
50828/2022-001AB & STARLINK-3290 & 547.9 & 3099.60 & 7 & 2.2 & 94.8 & 51.6 &      &      &      \\
50836/2022-001AK & STARLINK-3330 & 548.8 & 3578.54 & 4 & 2.1 &      & 45.8 &      &      &      \\
50839/2022-001AN & STARLINK-3320 & 548.6 & 3492.75 & 6 & 2.2 & 73.8 & 51.5 &      &      &      \\
50840/2022-001AP & STARLINK-3295 & 547.6 & 3162.72 & 10 & 2.1 & 102.3 & 37.8 & 6.8 & 1.0 &      \\
50842/2022-001AR & STARLINK-3302 & 546.7 & 2832.29 & 11 & 3.4 & 109.4 & 84.8 &      &      &      \\
50846/2022-001AV & STARLINK-3308 & 546.1 & 2502.19 & 12 & 2.6 & 70.9 & 58.6 &      & 0.5 &      \\
50849/2022-001AY & STARLINK-3311 & 545.5 & 2171.94 & 4 & 2.6 &      & 35.6 &      &      &      \\
52669/2022-053P & STARLINK-3972 & 546.4 & 477.68 & 5 & 1.5 & 101.5 & 30.8 &      & 0.8 &      \\
52671/2022-053R & STARLINK-4071 & 547.2 & 148.46 & 11 & 1.5 & 104.7 & 34.3 & 12.1 & 0.5 & 1.0 \\
53132/2022-083A & STARLINK-4063 & 547.2 & 2097.51 & 6 & 2.7 & 114.6 & 39.7 & 16.1 & 0.9 & 0.8 \\
53137/2022-083F & STARLINK-4145 & 547.7 & 2427.57 & 1 & 1.7 & 66.2 &      &      &      &      \\
53139/2022-083H & STARLINK-4185 & 546.3 & 1436.86 & 11 & 2.1 & 122.2 & 28.1 & 24.1 &      & 1.0 \\
53140/2022-083J & STARLINK-4170 & 546.7 & 1767.16 & 11 & 2.8 & 92.7 & 70.2 & 13.0 &      & 0.8 \\
53142/2022-083L & STARLINK-4101 & 546.0 & 861.97 & 1 & 2.5 & 75.3 &      &      &      &      \\
53151/2022-083V & STARLINK-4040 & 546.6 & 2278.75 & 11 & 1.2 & 98.7 & 85.0 &      & 0.8 &      \\
53152/2022-083W & STARLINK-4045 & 546.4 & 2939.61 & 8 & 1.9 & 106.9 & 28.1 &      &      & 0.5 \\
53153/2022-083X & STARLINK-4098 & 547.0 & 1948.56 & 7 & 2.3 & 96.1 &      &      &      &      \\
53154/2022-083Y & STARLINK-4077 & 546.4 & 2609.52 & 11 & 1.7 & 155.0 & 36.1 & 11.3 & 1.4 & 0.7 \\
53158/2022-083AC & STARLINK-4054 & 547.4 & 1617.95 & 1 & 1.9 &      &      &      &      & 1.4 \\
53170/2022-083AQ & STARLINK-4095 & 547.0 & 883.27 & 11 & 1.5 & 114.8 & 30.4 &      &      &      \\
53172/2022-083AS & STARLINK-4299 & 548.7 & 223.01 & 1 & 2.6 & 97.0 &      &      &      &      \\
53684/2022-105AN & STARLINK-4624 & 571.0 & 3551.23 & 2 & 1.8 & 75.4 & 33.2 &      &      &      \\
\hline
\end{tabular}
\tablefoot{Satellite versions v1.0 and v1.5 are as indicated. The column descriptions are otherwise identical to those in Table\,\ref{tab:satellite_passes_lba}.}
\end{table*}

\begin{table*}[!h]
\tiny
\centering
\caption{Second-generation Starlink satellites that have been detected between 110 and 188\,MHz. }
\label{tab:satellite_passes_hba_gen2}
\begin{tabular}{ll|rrr|ccc}
\hline
NORAD / COSPAR & Name & $d$ & $t_\mathrm{mid}$ & $n_\mathrm{TAB}$ & \multicolumn{3}{c}{$S_\nu$ (Jy)} \\
& &  (km) & (s) & & 116--124  & 150.05--153 & 157--165 \\
& &       &     & & (MHz) & (MHz) & (MHz) \\
\hline
58028/2023-156A & STARLINK-30514 & 489.3 & 3233.81 & 11 & 12.6 & 10.8 & 10.4 \\
58029/2023-156B & STARLINK-30518 & 488.2 & 2814.52 & 11 & 16.0 & 9.6 & 20.2 \\
58030/2023-156C & STARLINK-30549 & 489.1 & 2779.45 & 10 & 10.6 & 12.4 & 24.8 \\
58531/2023-192A & STARLINK-31012 & 488.0 & 845.05 & 12 & 3.6 & 59.4 & 76.9 \\
58535/2023-192E & STARLINK-31032 & 489.2 & 425.75 & 11 & 6.2 & 60.7 & 93.4 \\
58541/2023-192L & STARLINK-31045 & 558.7 & 1663.61 & 10 & 3.2 & 23.8 & 39.7 \\
59000/2024-036C & STARLINK-31529 & 490.2 & 3575.53 & 5 & 3.5 & 37.9 & 34.4 \\
59003/2024-036F & STARLINK-31429 & 489.4 & 1898.36 & 1 & 3.6 & 49.3 & 43.0 \\
59009/2024-036M & STARLINK-31488 & 489.1 & 1868.27 & 1 & 1.8 & 52.3 & 60.1 \\
59010/2024-036N & STARLINK-31528 & 481.4 & 1331.45 & 11 & 6.7 & 37.1 & 57.1 \\
59011/2024-036P & STARLINK-31491 & 482.3 & 1425.33 & 10 & 6.6 & 41.3 & 60.7 \\
59012/2024-036Q & STARLINK-31465 & 482.4 & 951.62 & 12 & 4.8 & 40.6 & 80.7 \\
59013/2024-036R & STARLINK-31389 & 488.1 & 1029.53 & 11 & 5.0 & 62.6 & 64.9 \\
59014/2024-036S & STARLINK-31454 & 486.9 & 639.58 & 11 & 105.6 & 37.5 & 96.0 \\
59015/2024-036T & STARLINK-31521 & 481.9 & 572.74 & 8 & 9.1 & 4.0 & 4.6 \\
59016/2024-036U & STARLINK-31526 & 453.2 & 1143.53 & 8 & 6.1 & 39.0 & 69.7 \\
59962/2024-107T & STARLINK-31980 & 454.3 & 1572.30 & 6 & 10.5 & 45.7 & 72.8 \\
59964/2024-107V & STARLINK-31979 & 453.7 & 2018.05 & 11 & 7.6 & 50.8 & 76.9 \\
60017/2024-111B & STARLINK-31987 & 454.4 & 2134.08 & 6 & 5.8 & 50.7 & 80.8 \\
60024/2024-111J & STARLINK-31990 & 453.8 & 1747.35 & 11 & 6.5 & 59.5 & 93.8 \\
60029/2024-111P & STARLINK-32006 & 454.5 & 2163.51 & 4 & 5.0 & 51.7 & 70.8 \\
60032/2024-111S & STARLINK-32017 & 453.2 & 1301.68 & 11 & 8.8 & 17.1 & 20.7 \\
60033/2024-111T & STARLINK-31998 & 453.7 & 1717.76 & 11 & 5.0 & 49.6 & 93.8 \\
60039/2024-112B & STARLINK-11140 [DTC] & 348.3 & 1203.93 & 3 & 20.2 & 10.3 & 7.8 \\
60040/2024-112C & STARLINK-11122 [DTC] & 346.9 & 1010.02 & 10 & 40.0 & 14.8 & 9.5 \\
60041/2024-112D & STARLINK-11149 [DTC] & 346.6 & 827.23 & 11 & 42.5 & 19.5 & 11.2 \\
60042/2024-112E & STARLINK-11120 [DTC] & 345.4 & 635.82 & 11 & 15.7 & 84.9 &      \\
60043/2024-112F & STARLINK-11086 [DTC] & 347.0 & 451.20 & 6 & 29.2 & 5.3 & 3.2 \\
60048/2024-112L & STARLINK-11135 [DTC] & 348.5 & 943.22 & 5 & 81.4 & 8.3 & 7.3 \\
60054/2024-112S & STARLINK-31735 & 454.3 & 2584.05 & 7 & 10.5 & 43.5 & 71.5 \\
\hline
\end{tabular}
\tablefoot{Satellites with [DTC] in their name are of the v2-Mini direct-to-cell (DTC) version, all other satellites are of the v2-Mini version. The column descriptions are otherwise identical to those of Table\,\ref{tab:satellite_passes_lba}. The passage mid-point times $t_\mathrm{mid}$ are with respect to the observation start time of July 19, 2024, at 07:30\,UTC.}
\end{table*}

\end{appendix}

\end{document}